# برهمکنش‌های فوق ریز در بلور $USb_2$

آرش فتحی[۱و۲]، سعید جلالی اسد آبادی[۳و۲] و محمد گشتاسبی راد[۱]

۱. گروه فیزیک، دانشگاه سیستان و بلوچستان، زاهدان
۲. گروه فیزیک، دانشگاه اصفهان، اصفهان
۳. مرکز پژوهش دانش و فناوری نانو، دانشگاه اصفهان، اصفهان

### چکیده
برهمکنش‌های فوق ریز در مکان اورانیوم از ترکیب آنتی فرومغناطیس $USb_2$ با استفاده از نظریه تابعی چگالی و روش امواج تخت بهبود یافته به علاوه اوربیتال‌های موضعی ($APW + lo$) بررسی شده‌اند. میزان جایگزیدگی الکترون‌های $f$ در این ترکیب مورد بررسی قرار گرفته است. نتایج نشان می‌دهند که الکترون‌های $5f$ تمایل دارند به خوبی با الکترون‌های رسانش هیبرید شوند.

واژه‌های کلیدی: ؟؟؟؟؟؟؟؟؟؟؟؟؟؟؟

## ۱. مقدمه

مطالعه ترکیبات دارای الکترون $f$ به خاطر رفتار چندگانه این الکترون‌ها از دیرباز تا کنون مورد بحث بوده است [۱]. همان گونه که نمی‌توان برای شعاع اتمی (یونی) اتم‌ها مقادیر ثابت یا معینی ارایه کرد، جایگزیدگی الکترون‌های $f$ نیز به محیط‌های بلوری بستگی دارد و از بلوری به بلور دیگر تغییر می‌کند [۲].

خواص یک بلور به میزان جایگزیدگی الکترون‌های آن بستگی دارد [۳] و چون این میزان جایگزیدگی معلوم نیست باید تدابیری اندیشیده شود که به وسیله آنها بتوان نسبت به میزان جایگزیدگی الکترون‌ها اطلاع کافی کسب کرد [۴ و ۵]. رفتار الکترون‌های $f$ در بسیاری از موارد در ترکیبات اورانیوم در مقایسه با ترکیبات عناصر خاکی متفاوت است [۶].

تحقیقات مختلفی که درباره بررسی رفتار الکترون‌های $f$ صورت گرفته همگی نتایج یکسانی را در مورد جایگزیدگی الکترون‌های $f$ ارایه نکرده‌اند. آوکی و همکاران با آزمایش‌های "دی هاس - ون آلفن" و "شابنیکوف - دی هاس" جرم سیکلوترونی را برای ترکیبات $UP_2$، $UAs_2$، $USb_2$ و $UBi_2$ به دست آورده‌اند، که با توجه به مقادیر بزرگ آنها الکترون‌های $5f$ را به صورت تقریبا غیر جایگزیده با سهمی در نوار رسانش معرفی کرده‌اند [۶]. ایشان همچنین خاطر نشان کرده‌اند که خصوصیت شبه دو بعدی سطوح فرمی در این ترکیبات پاد فرومغناطیسی عمدتا ناشی از وجود الکترون‌های رسانش در صفحات فرومغناطیسی اورانیوم شامل الکترون‌های $5f$ می‌باشد [۶]. این نکته بر آن دلالت دارد که الکترون‌های $5f$ در این ترکیبات تمایل دارند که با سایر الکترون‌ها در ناحیه ظرفیت هیبرید شوند که خود گویای خاصیت غیر جایگزیده بودن آنهاست.

از سوی دیگر آمورتی و همکاران با پیش فرض جایگزیدگی الکترونهای 5f بر اساس مدل میدان بلوری و تقریب میدان ملکولی با محاسبه آنتروپی مغناطیسی بر حسب شکافتگی میدان بلوری خواص مغناطیسی ترکیبات $UX_2$ (X=P, As, Sb) را تعبیر کرده‌اند [۷]. فرض جایگزیدگی ایشان از مقادیر بزرگ تجربی گشتاورهای مغناطیسی الکترونهای $f$ ۵ و به دنبال آن نسبت دادن ظرفیت ۴+ به یونهای اورانیوم ناشی شده است.

یکی از ابزارها و روشهای بررسی اتمها در یک ساختار بلوری از لحاظ انواع برهمکنش‌های الکترون - هسته، بررسی برهمکنش‌های فوق ریز از جمله گرادیان میدان الکتریکی ($EFG$) و میدان مغناطیسی فوق ریز ($HFF$) می‌باشد. با آشکارسازی همزمان $EFG$ و $HFF$ سر نخی از چگونگی وضعیت مغناطیسی و الکتریکی در نزدیکی هسته به دست می‌آید. محاسبات ساختار نواری ابتدا به ساکن نقش مهمی در تصحیح، رد یا قبول تفسیر داده‌های تجربی ایفا می‌کنند [۸]. تسوتسی و همکاران [۹] با استفاده از روش تجربی طیف سنجی موسبائر مقادیر $EFG$ و $HFF$ را برای سری $UX_2$ اندازه‌گیری کرده‌اند. یکی از برهمکنش‌های فوق ریز برهمکنش چهار قطبی هسته‌ای $\left(e^2Qq\right)$ است که در آن $e$، $Q$ و $q(V_{zz})$ به ترتیب عبارتند از بار الکترون، ممان چهار قطبی هسته و مؤلفه اصلی موازی با $EFG$. از آنجایی که این برهمکنش ناشی از برهمکنش بین گشتاورهای چهار قطبی الکتریکی و هسته‌ای است، مقدار آن برای تعیین توزیع گشتاور چهار قطبی الکتریکی در دستگاه‌های $f$ الکترونی و تعیین تقارن موضعی حول اتمهای مورد بررسی مفید واقع می‌شود [۹].

در این مقاله محاسبات به روش $(APW+lo)$ [۱۰] ابتدا در حالت آنتی فرومغناطیس در غیاب برهمکنش اسپین مدار انجام شده است. سپس برهمکنش نسبیتی اسپین - مدار را به روش وردش مرتبه دوم به حالت آنتی فرومغناطیس اضافه کرده‌ایم (که با استفاده از تقریب $GGA$ انجام شده) و در پایان به منظور در نظر گرفتن همبستگی‌های موجود بین الکترونهای $f$ تقریب $LDA+U$ را روی برهمکنش‌های قبلی اضافه کرده‌ایم. مزیت استفاده از هر تقریب روی تقریبهای دیگر در این است که در هر مرحله هامیلتونی جدید به عنوان یک اختلال به هامیلتونی قبلی (که در مرحله قبل قطری شده) اضافه می‌گردد و بدین ترتیب نیازی به قطری سازی کامل هامیلتونی جدید نیست. این کار در کاهش زمان محاسبات بسیار مؤثر است [۱۰].

با بررسی چگالی حالتها ($DOS$) برای ترکیب آنتی - فرو مغناطیسی $USb_2$، رفتار جایگزیدگی الکترونهای $f$ ۵ اتم اورانیوم را مورد بحث قرار داده‌ایم. همچنین مقادیر گرادیان میدان الکتریکی و میدان مغناطیسی فوق ریز را برای این ترکیب محاسبه کرده‌ایم.

## ۲. ساختار بلوری

سری بلورهای اورانیوم $UX_2$ $(X=Bi, Sb, As, P)$ دارای ساختار تتراگونال از نوع $Cu_2-Sb$ با گروه فضایی ($P4/nmm$ یا $D_{4h}^7$) و دمای نیل نسبتاً بالا $۱۸۰-۲۷۰°K$ هستند [۱۱ و ۱۲]. گشتاورهای مغناطیسی یونهای $U$ در بلور $USb_2$ درون صفحات (۰۰۱) به صورت فرومغناطیسی ردیف شده‌اند و این صفحات در راستای [۰۰۱] در یک چینش آنتی فرومغناطیس به صورت ($\uparrow\downarrow\downarrow\uparrow$) پشت سر هم قرار گرفته‌اند [۱۳]. یاخته واحد مغناطیسی در ترکیب $USb_2$ در امتداد جهت [۰۰۱] نسبت به یاخته شیمیایی دو برابر کشیده شده است (شکل ۱) [۶]. داده‌های بلور شناسی این ترکیبات در جدول ۱ آمده است [۷]. از نقطه نظر بلور شناسی همه اتمهای اورانیوم معادل هستند، اما دو جایگاه شبکه‌ای غیر یکسان برای قرارگیری اتمهای $X(I)$ و $X(II)$ وجود دارند. یاخته واحد با انتخاب مرکز وارون به عنوان مبدا دستگاه مختصات شامل دو اتم اورانیوم در $\left(\frac{1}{4}, \frac{1}{4}, u\right)$ و $\left(\frac{3}{4}, \frac{3}{4}, \bar{u}\right)$ و دو اتم $X(I)$ در $\left(\frac{3}{4}, \frac{1}{4}, 0\right)$ و $\left(\frac{1}{4}, \frac{3}{4}, 0\right)$ و دو اتم $X(II)$ در $\left(\frac{1}{4}, \frac{1}{4}, v\right)$ و $\left(\frac{3}{4}, \frac{3}{4}, \bar{v}\right)$ می‌باشد، که در آنها $u$ و $v$ پارامترهای داخلی ارایه شده در جدول ۱ هستند. اتمهای $X(I)$ و $X(II)$ از لحاظ الکتریکی و همچنین از لحاظ شعاع اتمی با هم متفاوت هستند [۷].

علی‌رغم متفاوت بودن گروههای فضایی در سری‌های $UX_2$ در تمام گونه‌های $UX_2$ تقارن موضعی اتم اورانیوم همیشه به

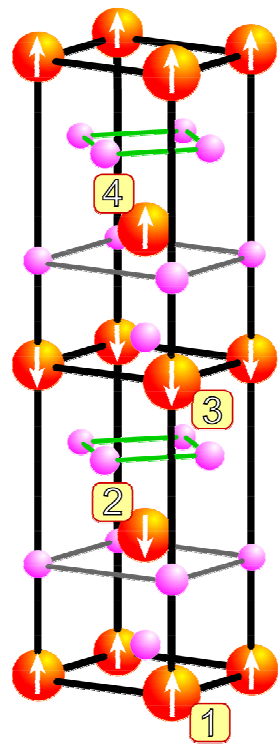

**شکل ۱.** یاخته واحد مغناطیسی ترکیب $USb_2$ با گروه فضایی $P4/mmm$.

صورت ساختار تتراگونال با گروه نقطه‌ای $C_{4V}$ است. نتایجی که آوکی و همکاران با انجام آزمایشات "دی هاس ون آلفن" به دست آورده‌اند حاکی از آن است که $USb_2$ دارای رفتار فلزی است [۶].

## ۳. نحوه محاسبه جمله تبادلی - همبستگی

جمله انرژی تبادلی - همبستگی موجود در معادلات تک ذره کان - شم به دلیل پیچیدگی در برهمکنش‌های کوانتومی موجود در دستگاه به طور دقیق قابل محاسبه نیست. به همین منظور برای محاسبه این بخش از انرژی مجبوریم از تقریبهایی استفاده کنیم. یکی از ساده‌ترین تقریبهایی که در نظریه تابعی چگالی استفاده می‌شود تقریب چگالی موضعی ($LDA$) است [۱۴]. کاربرد این تقریب در شرایطی مجاز و دارای دقت نسبتاً بالایی است که تغییرات نسبی چگالی دستگاه انحراف قابل توجهی نسبت به دستگاه همگن نداشته باشد. بخش تبادلی - همبستگی در بسیاری از حالات در انرژی کل دستگاه موثر است زیرا این

بخش در نواحی مختلف نه تنها به چگالی موضعی در آن ناحیه، بلکه به چگالی در نواحی مجاور نیز وابسته است. به دلیل این وابستگی با اضافه کردن جملات گرادیانی چگالی الکترونی به انرژی همبستگی تبادلی یا پتانسیل مربوط در تقریب $LDA$، تقریب دیگری به نام تقریب گرادیان تعمیم یافته $GGA$ [۱۴] وارد می‌شود. در این تقریب، گرادیان چگالی نقش وابستگی بخش تبادلی - همبستگی به چگالی نواحی مجاور را بازی می‌کند.

با توجه به موفقیتهای مدل چند نواری هابارد در توصیف دستگاههایی با الکترونهای قویاً همبسته که مدلهای $GGA$ و $LDA$ در مورد آنها پاسخگو نیست تقریب $LDA+U$ در سال ۱۹۹۱ توسط آنیسیمو ارایه شد [۱۵]. هدف اصلی تقریب $LDA+U$ کنترل جایگزیدگی اربیتالهای ترکیبات به طور قوی همبسته با استفاده از پارامتر $U$ هابارد به جای قوانین هوند است [۱۵]. انرژی کل حاصل از تقریب $LDA+U$ را می‌توان به شکل زیر نوشت [۱۵]:

$$E^{LDA+U} = E^{LSDA} + E^U - E^{dc} \qquad (1)$$

که در آن جمله اول انرژی کل حاصل از تقریب $LSDA$ است. جملات دوم و سوم به ترتیب انرژیهای حاصل از برهمکنش کولنی الکترونهای اربیتال همبسته بر طبق مدل هابارد و انرژی برهمکنش کولنی بر طبق مدل گاز همگن می‌باشند. روشی که توسط آنیسیمو و همکاران [۱۵] در سال ۱۹۹۱ اریه شده است به نام روش $HFM$ شهرت دارد که یکی از روشهای اعمال تقریب $LDA+U$ به شمار می‌رود. در این روش از پتانسیل تبادلی همبستگی ($LDA$) بدون در نظر گرفتن قطبیدگی اسپینی استفاده می‌شود. ما در اینجا به دلیل وجود یک دستگاه قطبیده اسپینی نمی‌توانیم از این روش استفاده کنیم. روش دیگری که برای این تقریب وجود دارد روشی موسوم به $AMF$ است که بیشتر برای دستگاههای فلزی و دستگاههای با همبستگی کم مناسب می‌باشد [۱۶]. ترکیب $USb_2$ علی‌رغم داشتن خصوصیت فلزی به عنوان یک دستگاه با بستگی زیاد شناخته شده است، بنابراین از این روش نیز صرفنظر می‌کنیم. روش شناخته شده بعدی در کاربرد تقریب $LDA+U$ روش $SIC$

جدول ۱. ثابتهای شبکه و پارامترهای داخلی بلورهای UX2

| ترکیب | a (A°) | C (A°) | c/a | پارامترهای داخلی | |
|---|---|---|---|---|---|
| | | | | u | V |
| UP₂ | ۳/۸۱۰ | ۷/۷۶۴ | ۲/۰۳۸ | ۰/۲۸۰ | ۰/۶۳۵ |
| UAs₂ | ۳/۹۵۴ | ۸/۱۱۶ | ۲/۰۵۳ | ۰/۲۸۰ | ۰/۶۳۵ |
| USb₂ | ۴/۲۷۰ | ۸/۸۸۴ | ۲/۰۴۸ | ۰/۲۸۰ | ۰/۶۳۵ |
| UBi₂ | ۴/۴۴۵ | ۸/۹۰۸ | ۲/۰۰۴ | ۰/۲۸۰ | ۰/۶۴۳ |

است که توسط آنیسیمو و همکاران در سال ۱۹۹۳ ارایه گردید [۱۷]. در اینجا یک روش تقریبی برای تصحیح خود برهمکنش (SIC) در نظر گرفته شده است. برای دستگاههای قوی همبسته و روش پتانسیل کامل این تقریب مناسبترین روش است که با توجه به توضیحات قبلی، ما در محاسبات خود از این روش استفاده کرده‌ایم. در تقریب $LDA+U$ (SIC) سه پارامتر وجود دارند که انرژی کل دستگاه را می‌توان با آنها تنظیم کرد. این سه پارامتر عبارتند از پارامتر کولنی $U$، پارامتر تبادلی استتار شده $J$ و ماتریس چگالی $n^{\sigma}_{mm'}$. تغییرات ناشی از تغییر پارامتر تبادلی کوچک و قابل اغماض است؛ بنابراین در اینجا پارامتر تبادلی ثابت و برابر مقدار تجربی $J = 0/51\ eV$ اختیار شده است. همچنین $U = 4/5\ eV$ انتخاب شده است [۱۸]. در این روش پیشنهاد می‌شود از یک پتانسیل موثر $U_{eff} = U - J$ با $J = 0$ استفاده شود [۱۷]. که در اینجا ما نیز این پیشنهاد را به کار برده و از مقادیر $U_{eff} = 3/99\ eV$ و $J = 0$ استفاده کرده‌ایم. انتگرال کولمبی U به عنوان پارامتر ورودی در محاسبات $LDA+U$ از یک سو به دستگاه مورد بررسی بستگی دارد، و می‌تواند برای یک یک اتم (در اینجا اتم $U$) از یک ترکیب به ترکیبات دیگر تغییر کند. از سوی دیگر مقدار این پارامتر به نوع محاسبات نیز بستگی دارد و مثلاً برای آن مقادیر متفاوتی از روشهای $LAPW$ و $LMTO$ حاصل می‌شود.

در اینجا از مقدار $U = 4/5 eV$ مربوط به اتم اورانیوم استفاده شده است لذا نتایج گزارش شده در این مقاله خطای ناشی از لحاظ نکردن مقدار دقیق پارامتر هابارد را در بر دارند.

## ۴. جزئیات محاسبات

ما در این مقاله از کد نرم افزاری $WIEN2k$ [۱۰] و از روش امواج تخت بهبود یافته به علاوه اربیتالهای موضعی [۱۹] $(APW+lo)$ از طریق یک حل خودسازگار، در یک فرایند تکرار شونده استفاده می‌کنیم. این برنامه بر نظریه تابعی چگالی $(DFT)$ استوار است و به حل معادلات کان – شم [۲۰] می‌پردازد به صورتی که با انتخاب کره‌های مناسب موفین – تین حول هر یک از اتمها، فضای درون هر یاخته به دو ناحیه تقسیم می‌شود. تابع چگالی، تابعی پتانسیل و توابع موج الکترونهای ظرفیت در درون کره‌ها برحسب هارمونیکهای شبکه و در خارج از آنها بر حسب امواج تختی که بردار موج آنها از تقارنهای گروه فضایی بلور تبعیت می‌کنند بسط داده می‌شوند.

به منظور همگرایی سریعتر ابتدا معادله نسبیتی دیراک – فوک برای کلیه الکترونهای هر یک از اتمهای غیر معادل در ترکیب بلوری حل و یک چگالی اولیه برای ساختن پتانسیل کان – شم معرفی می‌شود. به منظور دقت بیشتر، معادله نسبیتی دیراک – فوک مجددا برای الکترونهای مغزه در هر مرحله از حل خود سازگار، محاسبه می‌شود.

با رعایت شرط عدم همپوشانی کرات موفین – تین مقادیر $R_{MT} = 2/2\ a.u.$ و $R_{MT} = 2/8\ a.u.$ به ترتیب برای $Sb$ و

$U$ انتخاب شده‌اند. بیشینه عدد کوانتومی مداری برای توابع موج در داخل کرات اتمی برابر $l_{max}=۱۰$ قرار داده شده است. بردار موج قطع برای بسط تابع موج بر حسب امواج تخت در ناحیه بین جایگاهی برابر با $K_{max}=\frac{۷}{R_{MT}^{min}}$ پس از تحقیق همگرایی مطلوب نتایج انتخاب شده است که در آن $R_{MT}^{min}$ کوچکترین شعاع موفین ـ تین در یاخته واحد است. چگالی بار و پتانسیل در ناحیه بین جایگاهی تا $G_{max}=۱۴$ بسط داده شده‌اند. یک مش متشکل از تعداد ۱۰۸ نقطه k خاص برای محاسبه انتگرال حالتهای ظرفیت در لبه ناحیه کاهش ناپذیر بریلوین که متناظر با شبکه ۳×۱۵×۱۵ است بر طبق روش مونخارست ـ پک در نظر گرفته می‌شود [۲۱]. با رسم نمودار انرژی بر حسب تعداد نقطه‌های فضای وارون مشاهده می‌شود که به ازای ۹۰۰ نقطه $K$ متناظر با ۱۰۸ نقطه خاص انرژی کل همگرا می‌شود.

جفت شدگی اسپین ـ مدار در همه محاسبات با استفاده از روش وردش مرتبه دوم و با یک انرژی قطع $E_{cut}^{so}=۳Ry$ اعمال شده است. از سو برای حالتهای $P۵$ در لانتانیدها اضافه کردن اربیتالهای موضعی نسبیتی ($RLO$) سبب کاهش تعداد پایه‌های مورد نیاز از طریق کاهش انرژی قطع مورد نیاز محاسبات اسپین- مدار، $E_{cut}^{so}$، می‌شود، که در نتیجه آن زمان محاسبات نیز کاهش می‌یابد. از سوی دیگر استفاده از اربیتالهای موضعی نسبیتی، به دلیل محدودیتهای محاسباتی فعلی موجود در کد $WIEN۲k$ مانع به دست آمدن مقادیر دقیق برای کمیتهایی که در حوالی هسته‌ها ($r=۰$) حایز اهمیت هستند مانند $HFF$ و $EFG$ می‌شود [۲۲]. بنابراین در اینجا ترجیح داده شد که محاسبات با صرف زمان بیشتر بدون استفاده از اربیتالهای موضعی نسبیتی به منظور افزایش دقت محاسبات انجام شوند.

محاسبات برای الکترونهای مغزه به طور کاملاً نسبیتی و برای الکترونهای ظرفیت هم به صورت نسبیتی ـ نرده‌ای انجام شده‌اند. هامیلتونی برهمکنش اسپین ـ مدار به صورت زیر تعریف می‌شود [۲۳]:

$$H_{SO}=\frac{\hbar}{۲Mc^۲}\frac{۱}{r}\frac{dV}{dr}\begin{bmatrix}\vec{\sigma}\vec{l} & ۰ \\ ۰ & ۰\end{bmatrix} \quad (۲)$$

واضح است که این برهمکنش با عدد کوانتومی مداری $l$ متناسب است. بنابراین برای عناصر سنگین جدول تناوبی که دارای اربیتالهای $d$ یا $f$ پرنشده هستند این برهمکنش مهم بوده و قابل صرفنظر کردن نیست. به همین دلیل ما در محاسبات خود در مورد ترکیب $USb_2$ اندرکنش اسپین ـ مدار را لحاظ کرده‌ایم.

## ۵. برهمکنش‌های فوق ریز

با در نظر گرفتن ساختار برای هسته اختلاف ترازهای اتمی از مرتبهٔ $\mu eV$ خواهد شد که آن را ساختار فوق ریز می‌نامند. اگر به هسته یک توزیع چند قطبی الکتریکی نسبت دهیم برهمکنش ابر الکترونی با چند قطبیهای هسته را برهمکنش الکتریکی فوق ریز می‌نامند که اولین جمله غیر صفر ناشی از برهمکنش گرادیان میدان الکتریکی با چهارقطبی الکتریکی هسته است. اگر به هسته یک توزیع چند قطبی مغناطیسی نسبت دهیم برهمکنش گشتاور مغناطیسی هسته با میدان مغناطیسی ناشی از حرکت الکترون در محل هسته را برهمکنش مغناطیسی فوق ریز می‌نامند.

### ۵. ۱. گرادیان میدان الکتریکی

تمام هسته‌هایی که عدد کوانتومی اسپینی آنها ($I$) بزرگتر یا مساوی یک باشد دارای توزیع بار هسته‌ای غیر کروی و ممان چهار قطبی هسته‌ای $Q$ می‌باشند [۲۴]. حال به طور خلاصه نحوهٔ محاسبهٔ گرادیان میدان الکتریکی و نیز چگونگی تعین مولفه اصلی گرادیان میدان الکتریکی ($V_{zz}$) را به روش محاسباتی بیان می‌کنیم.

گرادیان میدان الکتریکی یک تانسور متقارن مرتبه دو با رد صفر با پنج مولفه مستقل است که به عنوان مشتق مرتبه دوم پتانسیل ناشی از ابر الکترونی در اطراف هسته هر اتم تعریف می‌شود. این پارامتر مقیاسی از انحراف چگالی بار الکتریکی از

تقارن کروی حول هستهٔ مورد نظر است. در دستگاه مختصات محورهای اصلی تانسور EFG قطری بوده و عناصر قطری آن $V_{xx}, V_{yy}, V_{zz}$ در رابطهٔ $V_{xx}+V_{yy}+V_{zz}=0$ صدق می‌کنند (چون ردتانسور گرادیان میدان الکتریکی صفر است بنابراین مشخص کردن دو تای آنها کافی است) سه محور دستگاه محورهای اصلی طوری نامگذاری می‌شوند که $|V_{xx}|\leq|V_{yy}|\leq|V_{zz}|$ باشد. به جای تعیین دو آرایه از سه آرایه قطری تانسور EFG معمولاً زوج $\eta$ و $V_{zz}$ مشخص می‌شوند که پارامتر عدم تقارن $\eta$ به شکل زیر تعریف می‌شود:

$$\eta = \frac{V_{xx}-V_{yy}}{V_{zz}}. \qquad (۳)$$

پارامتر عدم تقارن در مورد ترکیب مورد بحث در این پژوهش در مکان اتم اورانیوم بنابر تقارن نقطه‌ای آن برابر با صفر است و بنابراین تنها پارامتری که در این رهیافت مورد بررسی قرار خواهد گرفت $V_{zz}$ یا اندازه گرادیان میدان الکتریکی در مکان هسته اورانیوم است [۸]. بخش عمدهٔ EFG ناشی از ناهمسانگردی توزیع بار نزدیک هسته‌هاست. گرادیان میدان الکتریکی دارای دو سهم به نامهای شبکه و ظرفیت می‌باشد که به صورت زیر محاسبه می‌شوند [۲۵]:

$$V_{zz}=\left[\frac{۵}{\pi}\right]^{\frac{۱}{۲}}\lim\left[\frac{V_{۲۰}}{r^۲}\right], \qquad (۴)$$

که در آن $V_{۲۰}$ مؤلفهٔ $M=۲$ و $L=۰$ بسط پتانسیل در مکان هسته است که از رابطهٔ زیر به دست می‌آید:

$$V_{۲۰}(r=۰)=-۲\sqrt{\frac{۴\pi}{۵}}\int_۰^{R_t}\frac{\rho_{۲۰}}{r}dr$$
$$+۲\sqrt{\frac{۴\pi}{۵}}\int_۰^{R_t}\frac{\rho_{۲۰}}{r}(\frac{r}{R})^۵dr \qquad (۵)$$
$$+۲\sqrt{\frac{۴\pi}{۵}}\frac{۵}{R^۲}\sum_{\vec{k}}V(\vec{k})j_۲(kR_t)Y_{۲۰}(\hat{k}),$$

که در آن $\rho_{۲۰}$، $Y_{۲۰}$، $j_۲$ و $R_t$ به ترتیب مؤلفهٔ $M=۲$ و $L=۰$ بسط چگالی، یک هماهنگ کروی ($L=۰, M=۲$) تابع بسل کروی ($L=۲$) و شعاع کرهٔ موفین - تین است. جملهٔ اول در معادلهٔ (۵) ناشی از چگالی الکترونی غیر کروی الکترونهای ظرفیت (و شبه مغزه) درون کره‌های موفین - تین است که به

آن EFG ظرفیت گویند و مجموع دو جملهٔ آخر که از بسط چند قطبی‌ها به وجود آمده‌اند در قیاس با مدل قدیمی بار نقطه‌ای سهم شبکه در تولید گرادیان میدان الکتریکی نامیده می‌شود. این سهم‌ها را می‌توان به صورت سهم‌های جزئی‌تر بر طبق سهم‌های متفاوتی از توابع موج که $s-s$، $d-d$ و $p-p$ نامیده می‌شوند بیان کرد. حال تابع عدم تقارن اربیتال‌های $p$ و $d$ را به صورت زیر تعریف می‌کنیم:

$$\Delta p = \frac{۱}{۲}(p_x+p_y)-p_z \qquad (۶)$$

$$\Delta d = (d_{xy}+d_{x^۲-y^۲})-\frac{۱}{۲}(d_{xz}+d_{yz})-d_{z^۲} \qquad (۷)$$

$\Delta p$ و $\Delta d$ بیانگر انحراف از حالت تقارن کروی برای بارهای اربیتال‌های $p$ و $d$ در داخل کره موفین - تین و $p_i$ و $d_i$ به ترتیب بیانگر تعداد الکترون‌های مربوط به اربیتال‌های $p$ و $d$ درون کره موفین - تین تا انرژی فرمی است. هر چه مقدار این عدم تقارن توزیع بار در یک اربیتال بیشتر باشد سهم آن اربیتال در گرادیان میدان الکتریکی بیشتر است. ما این محاسبات را به روش‌های $GGA$ و $GGA+SO$ و همچنین با در نظر گرفتن تقریب $LDA+U$ انجام داده و نتایج را با یکدیگر مقایسه کرده‌ایم.

### ۵. ۲. میدان مغناطیسی فوق ریز

با اندازه‌گیری میدان‌های فوق ریز اطلاعاتی در مورد خواص الکترونی جامدات حاصل می‌شود که با روش‌های دیگر قابل دستیابی نیستند. اما عموماً تفسیر مقادیر اندازه‌گیری شده سر راست و آسان نیست. تمامی مطالعات نظری حاصل از محاسبات ابتدا به ساکن دستگاه‌های بس ذره‌ای بر اساس نظریهٔ تابعی چگالی اسپینی با روش گرادیان شیب تعمیم یافته $GGA$ به عنوان یک ابزار تکمیل کننده در بررسی میدان‌های فوق ریز محسوب می‌شوند [۸].

میدان فوق ریز را میدان مغناطیسی ناشی از حرکت مداری و اسپینی الکترون‌ها در حوالی هسته تعریف می‌کنیم. برای سادگی در محاسبات از ساختار چند قطبی‌های مغناطیسی که به هسته نسبت می‌دهیم تنها دو قطبی مغناطیسی را بر

**جدول ۲.** گرادیان میدان الکتریکی، سهم شبکه، سهم ظرفیت با استفاده از تقریبهای مختلف و در دمای $T=0^\circ K$ ($\times 10^{21} \frac{V}{m^2}$)

| T=۰ | GGA | GGA+SO | LDA(GGA+SO)+U | Other's work |
|---|---|---|---|---|
| Total | ۸/۷۰۲۲۱ | ۹/۰۸۹۲۲ | ۱۱/۱۳۵۶۹ | ۱۶±۴ |
| Valance | ۸/۷۲۳۷۰ | ۶/۱۳۸۱۰ | ۸/۳۶۱۱۱۵ | – |
| Lattice | −۰/۰۲۱۵ | ۲/۹۵۱۱۰ | ۲/۷۷۴۵۰۰ | – |

می‌گزینیم.

برهمکنش بین میدان مغناطیسی هسته و گشتاور دو قطبی اسپینی الکترون (سهم/اسپینی) در هر مکان غیر از هسته با رابطهٔ زیر داه می‌شود که آن را جملهٔ دو قطبی اسپینی نامند.

$$H_{HF}^d = \frac{\mu_0 \mu_B}{4\pi} \left( -\frac{\vec{\sigma}}{r^3} + \frac{3(\vec{\sigma}\cdot\vec{e}_r)\vec{e}_r}{r^3} \right) \cdot \vec{\mu}_I \quad (8)$$

حال می‌خواهیم میدان مغناطیسی حاصل از دو قطبی هسته‌ای را در مکان $r=0$ یعنی در درون هسته نیز محاسبه کنیم. این محاسبه برای آن انجام می‌شود که الکترونهای $s$ قدرت نفوذ در هسته را دارند. بنابراین انرژی برهمکنشی این الکترونها با میدان مغناطیسی حاصل از هسته (سهم تماس فرمی) نیز محاسبه می‌شود که به قرار زیر است

$$H_{HF}^{CF} = -\frac{2\mu_0}{3} \vec{\mu}_e \cdot \vec{\mu}_I \quad (9)$$

تا اینجا انرژی برهمکنشی میدان مغناطیسی هسته و گشتاور اسپینی الکترون را به دست آوردیم. در ادامه انرژی برهمکنشی این میدان و حرکت مداری الکترون (سهم اربیتالی) با استفاده از محاسبات اختلال مرتبه اول طبق رابطه زیر داده می‌شود.

$$\Delta H_{HF}^O = \frac{\mu_B \mu_0}{2\pi \hbar r^3} \vec{\mu}_I \cdot \vec{L}_i \quad (10)$$

بدین ترتیب هامیلتونی برهمکنش یک الکترون با میدان مغناطیسی $B_n$ ناشی از هسته را تعیین و با استفاده از آن میدان مغناطیسی فوق ریز را در حوالی هسته به صورت زیر خلاصه می‌کنیم: [۲۴]

$$H_{HF} = H_{HF}^{CF} + H_{HF}^O + H_{HF}^d \quad (11)$$

در این رابطه $H_{HF}^{CF}$ سهم تماس فرمی، $H_{HF}^O$ سهم اربیتالی و $H_{HF}^d$ سهم دو قطبی اسپینی هستند.

## ۶. بحث و نتیجه‌گیری

### ۶.۱. گرادیان میدان الکتریکی:

مقادیر گرادیان میدان الکتریکی، نزدیک هستهٔ اتمهای اورانیوم با استفاده از تقریبهای $GGA$ و $GGA+SO$ (با در نظر گرفتن اثر جفت شدگی اسپین مدار و بدون آن) و همچنین وارد کردن تقریب $LDA+U$ (در تمام محاسبات انجام شده منظور از $LDA+U$ در واقع $(GGA+SO)+U$ است که پارامتر هابارد $U$ به صورت تدریجی پس از همگرایی محاسبات $GGA+SO$ وارد شده است.) در جدول ۲ آورده شده‌اند و با مقادیر تجربی مقایسه شده‌اند. به علاوه گرادیانهای میدان الکتریکی در جدول ۲ به مولفه‌های آن ناشی از سهم شبکه و سهم ظرفیت مطابق با رابطه (۵) تجزیه شده است.

از آنجا که اتم اورانیوم یک عنصر سنگین $^{92}U$ است، انتظار داریم که برهمکنش اسپین ـ مدار بر آن مؤثر باشد. نتایج نشان می‌دهند که برهمکنش اسپین – مدار از یک سو باعث کاهش مقادیر $V_{zz}^{Val}$ و از سوی دیگر افزایش $V_{zz}^{Latt}$ می‌شوند، به طوری که در مجموع گرادیان میدان الکتریکی کل با مقادیر تجربی در توافق بیشتری قرار می‌گیرند.

همچنین درنظر گرفتن تقریب $LDA+U$ و وارد کردن پارامترهای موجود در هامیلتونی هابارد به طور موثر، $J_{effective} = 0$, $U_{effective} = U - J$، موجب تغییر در نتایج و بهبود آنها شده است. نتایج نشان می‌دهند که تاثیر $LDA+U$ بر $V_{zz}^{Val}$ بر خلاف اثر برهمکنش اسپین – مدار سبب افزایش مولفه ظرفیت گرادیان میدان الکتریکی در راستای توافق بیشتر با تجربه شده است. این در حالی است که نتایج مولفه شبکه

گرادیان میدان الکتریکی در اثر $LDA+U$ کماکان همچون اثر برهمکنش اسپین – مدار افزایش یافته به طوری که در مجموع به توافق چشمگیرتری منجر شده است. این بدان معنی است که توافق بیشتر با تجربه ناشی از مولفه ظرفیت گرادیان میدان الکتریکی است که به خوبی توسط $LDA+U$ بر خلاف $GGA+SO$ به سمت مقدار تجربی افزایش می‌یابد.

با توجه به تأثیر مثبت تدریجی اعمال $LDA+U$ بر نتایج حاصل از محاسبات $GGA+SO$ به نظر می‌رسد که الکترونهای اتمهای اورانیوم در این ماده تمایل بیشتری به جایگزیده بودن از خود نشان می‌دهند، که البته به این نکته در بخش چگالی حالتها باز خواهیم گشت.

سهم اربیتالهای مختلف در گرادیان میدان الکتریکی در بلور $USb_2$ با استفاده از تقریبهای $GGA$ و $GGA+SO$ و $LDA+U$ به ترتیب در جداول ۳ و ۴ و ۵ آورده شده است. نتایج برای گروه‌های نقطه‌ای مختلف اورانیوم ارایه شده‌اند، که با توجه به رفتار مشابه آنها توجه خود را به یکی از اتمها مثلاً اتم (۴) $U$ معطوف می‌کنیم. نتایج قبل از اعمال برهمکنش‌های اسپین ـ مدار و $LDA+U$ (نتایج $GGA$) نشان می‌دهند که سهم مشارکت $p-p$ (مولفه‌های $up$ و $dn$) در تولید مولفه ظرفیت گرادیان میدان الکتریکی ($V_{z_0}$) نسبت به سهم‌های مشارکت اربیتالهای $s$، $d$ و $f$ بیشتر است. اما پس از اعمال برهمکنش اسپین – مدار ملاحظه می‌شود که سهم مشارکت $f-f$ در حالت اسپین بالا (۱۹/۲۴۸) به مراتب بیشتر حتی از سهم مشارکتی $p-p$ می‌شود (۴/۶۲۳-). این در حالی است که هنوز اربیتال $f$ در حالت اسپین پایین تمایلی به مشارکت بیشتر از خود نشان نمی‌دهد (۱/۰۰۹-). نتیجه این تمایل حالت اسپین بالای اربیتال $f$ از یک سو و عدم تمایل حالت اسپین پایین از سوی دیگر در مجموع به گونه‌ای رفتار می‌کند که سهم مشارکت $f-f$ نسبت به وضعیت قبل از اعمال برهمکنش اسپین – مدار تفاوت زیادی را در نتیجه نهایی مولفه ظرفیت گرادیان میدان الکتریکی ایجاد نمی‌کند. عکس این وضعیت در مورد سهم‌های مشارکتی $p-p$ یعنی تمایل تقریباً یکسان حالتهای اسپین بالا و پایین آن سبب حفظ مشارکت بیشتر

اربیتال $p$ در مولفه $V_{z_0}$ می‌شود. جدول (۵) وضعیت مشابهی را برای نتایج $LDA+U$ نسبت به نتایج $GGA$ پیش بینی می‌کند، که در آن مجدداً تمایلهای یکسان اربیتال $p$ در حالتهای اسپین بالا و پایین و تمایلهای متفاوت اربیتال $f$ در حالتهای اسپین بالا و پایین دیده می‌شود.

یاخته شیمیایی تمام ترکیباتی که دارای ساختار $Cu_2-Sb$ هستند در شکل ۲ نمایش داده شده است. ترکیب $USb_2$ نیز به دلیل دارا بودن چنین ساختاری از یاخته شیمیایی مشابهی برخوردار است. اتمهای اورانیوم در این یاخته به لحاظ موقعیت فضایی کاملاً یکسان هستند. برای نیل به ساختار مغناطیسی و چینش صحیح اسپین‌ها باید اتمهای اورانیوم را از یکدیگر متمایز کرد که در این صورت گروه فضایی از $(P4/nmm)$ به $(P4/mm)$ تغییر می‌یابد. با توجه به تغییر گروه فضایی انتظار داریم که اتمهای (۱) و (۲) در شکل ۳ به لحاظ موقعیت فضایی، تفاوت اندکی با هم داشته باشند. بدین ترتیب می‌توان ادعا کرد که در یاخته واحد مغناطیسی که در شکل ۱ رسم شده است توزیع ابر الکترونی اطراف اتمهای (۱) و (۳) با هم مشابه باشند.

اما از مقایسه سهم‌های مختلف گرادیان میدان الکتریکی اتمهای اورانیوم مشاهده می‌شود که همخوانی بین سهم اتمهای (۱) و (۴) با هم و (۲) و (۳) نیز با هم مشابه می‌باشند و سهم گرادیان میدان الکتریکی اتم (۱) با اتمهای (۲) و (۳) متفاوت است. چنین استنباط می‌شود که رفتار متفاوت این اتمها ناشی از متفاوت بودن موقعیت فضایی آنها نیست، بلکه تفاوت عمده از چینش اسپینی خاص آنها و در کل نشأت گرفته از ساختار مغناطیسی بلور است. به دلیل هم جهت بودن اسپین اتمهای (۱) و (۴) علی رغم تفاوت در موقعیت فضایی، دارای سهم‌های $EFG$ نزدیکی هستند و وجود اسپین مخالف در اتمهای (۱) و (۴) نسبت به (۲) و (۳) باعث به وجود آمدن اختلاف در مقادیر سهم‌های گرادیان میدان الکتریکی آنها شده است. این در صورتی است که این اتمها به لحاظ موقعیت فضایی تقریباً با هم مشابه هستند. بنابراین می‌توان سهم عمده در تقارن توزیع بار حول اتمهای اورانیوم را به ساختار مغناطیسی ربط داد. به عبارت دیگر برطبق محاسبات ابتدا به ساکنی که ما انجام داده‌ایم

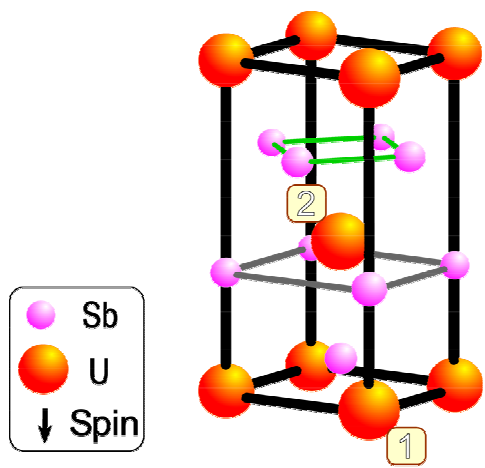
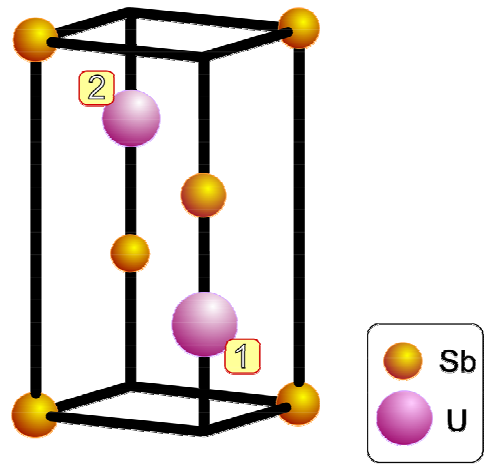

**شکل ۳.** یاخته واحد شیمیایی ترکیب $USb_2$ با گروه فضایی $P4/nmm$.

**شکل ۲.** یاخته شیمیایی ترکیبات $Cu_2-Sb$ با گروه فضایی $P4/nmm$.

**جدول ۳.** مقادیر سهم اربیتالهای مختلف در EFG با استفاده از تقریب $GGA$ ($\times 10^{21} \frac{V}{m^2}$).

| GGA | s-d (up) | s-d (dn) | p-p (up) | p-p (dn) | d-d (up) | d-d (dn) | p-f (up) | p-f (dn) | f-f (up) | f-f (dn) |
|---|---|---|---|---|---|---|---|---|---|---|
| U(۱) | -۰/۰۶۶ | ۰/۱۰۸ | ۲/۱۲۱ | ۴/۵۳۸ | ۰/۱۲۲ | -۰/۱۵۸ | ۰/۰۰۸ | -۰/۰۱۴ | ۱/۳۴ | -۰/۶۵ |
| U(۲) | ۰/۱۰۶ | -۰/۰۶۶ | ۴/۵۴ | ۲/۱۲۵ | -۰/۱۶۰ | ۰/۱۲۲ | -۰/۰۱۴ | ۰/۰۰۶ | -۰/۶۴۲ | ۱/۳۲۴ |
| U(۳) | ۰/۱۱۰ | -۰/۰۶۴ | ۴/۵۸۶ | ۲/۱۸۴ | -۰/۱۵۶ | ۰/۱۳۴ | -۰/۰۱۴- | ۰/۰۰۴ | -۰/۶۵۳ | ۱/۴۱۷ |
| U(۴) | -۰/۰۶۴ | ۰/۱۱۰ | ۲/۱۸۹ | ۴/۵۸۰ | ۰/۱۳۷ | -۰/۱۵۶ | ۰/۰۰۴ | -۰/۰۱۴ | ۱/۴۲۵ | -۰/۶۵۲ |

**جدول ۴.** مقادیر سهم اربیتالهای مختلف در EFG با استفاده از تقریب $GGA+SO$ ($\times 10^{21} \frac{V}{m^2}$).

| GGA+SO | s-d (up) | s-d (dn) | p-p (up) | p-p (dn) | d-d (up) | d-d (dn) | p-f (up) | p-f (dn) | f-f (up) | f-f (dn) |
|---|---|---|---|---|---|---|---|---|---|---|
| U(۱) | ۰/۰۳۰ | ۰/۲۵۰ | -۴/۶۲۷ | -۷/۶۰۹ | -۰/۰۴۳ | -۰/۰۸۶۱ | -۰/۰۲۲ | -۰/۰۴۲ | ۱۹/۲۲۵ | -۱/۰۱۱ |
| U(۲) | ۰/۲۵۰ | ۰/۰۳۰ | -۷/۵۹۳ | -۴/۶۱۵ | -۰/۰۸۶۰ | -۰/۰۴۲ | -۰/۰۴۲ | -۰/۰۲۲ | -۱/۰۱۰ | ۱۹/۲۴۹ |
| U(۳) | ۰/۲۵۰ | ۰/۰۳۰ | -۷/۵۹۳ | -۴/۶۱۳ | -۰/۰۹۶۰ | -۰/۰۴۲ | -۰/۰۴۲ | -۰/۰۲۲ | -۱/۰۰۹ | ۱۹/۲۶۹ |
| U(۴) | ۰/۰۳۰ | ۰/۲۵۰ | -۴/۶۲۳ | -۷/۶۱۱ | -۰/۰۴۲ | -۰/۰۸۶۰ | -۰/۰۲۲ | -۰/۰۴۲ | ۱۹/۲۴۸ | -۱/۰۰۹ |

**جدول ۵.** مقادیر سهم اربیتالهای مختلف در $EFG$ با استفاده از تقریب $LDA+U$ ($\times 10^{21} \frac{V}{m^2}$).

| LDA+U | s-d (up) | s-d (dn) | p-p (up) | p-p (dn) | d-d (up) | d-d (dn) | p-f (up) | p-f (dn) | f-f (up) | f-f (dn) |
|---|---|---|---|---|---|---|---|---|---|---|
| U(۱) | ۰/۰۳۰ | ۰/۲۵۰ | -۴/۷۴۶ | -۷/۵۵۸ | -۰/۱۸۷ | -۰/۸۷۰ | -۰/۰۲۴ | -۰/۰۴۲ | ۲۱/۴۲۱ | -۱/۰۲۵ |
| U(۲) | ۰/۲۵۲ | ۰/۰۳۰ | -۷/۵۳۶ | -۴/۷۳۱ | -۰/۸۶۶ | -۰/۱۹۷ | -۰/۰۴۲ | -۰/۰۲۴ | -۱/۰۲۳ | ۲۱/۴۵۱ |
| U(۳) | ۰/۲۵۲ | ۰/۰۳۰ | -۷/۵۴۳ | -۴/۷۳۹ | -۰/۸۶۷ | -۰/۱۹۸ | -۰/۰۴۲ | -۰/۰۲۴ | -۱/۰۲۶ | ۲۱/۴۴۴ |
| U(۴) | ۰/۰۳۰ | ۰/۲۵۰ | -۴/۷۴۹ | -۷/۵۶۵ | -۰/۱۸۸ | -۰/۸۶۹ | -۰/۰۲۴ | -۰/۰۴۲ | ۲۱/۴۲۸ | -۱/۰۲۹ |

جدول ۶. توابع عدم تقارن اربیتالهای $d, P$ با استفاده از تقریب $GGA$.

| GGA | $\Delta p$ (up) | $\Delta p$ (dn) | $\Delta d$ (up) | $\Delta d$ (dn) | $\Delta p$ | $\Delta d$ |
|---|---|---|---|---|---|---|
| U(۱) | ۰/۹۵۷۹ | ۰/۹۵۶۵ | ۰/۰۹۹۶ | ۰/۰۸۸۴ | ۱/۹۱۴۴ | ۰/۱۸۸۰ |
| U(۲) | ۰/۹۵۶۵ | ۰/۹۵۷۹ | ۰/۰۸۸۵ | ۰/۰۹۹۵ | ۱/۹۱۴۴ | ۰/۱۸۸۰ |

جدول ۷. توابع عدم تقارن اربیتالهای $d, P$ با استفاده از تقریب $GGA+SO$.

| $GGA+SO$ | $\Delta p$ (up) | $\Delta p$ (dn) | $\Delta d$ (up) | $\Delta d$ (dn) | $\Delta p$ | $\Delta d$ |
|---|---|---|---|---|---|---|
| U(۱) | ۰/۹۵۳۴ | ۰/۹۵۰۵ | ۰/۱۱۶۰ | ۰/۱۰۹۲ | ۱/۹۰۳۹ | ۰/۲۲۵۲ |
| U(۲) | ۰/۹۵۰۵ | ۰/۹۵۳۴ | ۰/۱۰۹۲ | ۰/۱۱۶۰ | ۱/۹۰۳۹ | ۰/۲۲۵۲ |

جدول ۸. توابع عدم تقارن اربیتالهای $d, P$ با استفاده از تقریب $LDA+U$.

| $LDA+U$ | $\Delta p$ (up) | $\Delta p$ (dn) | $\Delta d$ (up) | $\Delta d$ (dn) | $\Delta p$ | $\Delta d$ |
|---|---|---|---|---|---|---|
| U(۱) | ۰/۹۵۱۹ | ۰/۹۴۷۱ | ۰/۱۰۹۲ | ۰/۰۹۱۱ | ۱/۸۹۹۰ | ۰/۲۰۰۳ |
| U(۲) | ۰/۹۴۷۱ | ۰/۹۵۱۹ | ۰/۰۹۰۹ | ۰/۱۰۹ | ۱/۸۹۹۰ | ۰/۱۹۹۹ |

مشخص شد که مهمترین عامل در ایجاد گرادیان میدان الکتریکی ساختار مغناطیسی و به عبارت دیگر همان چینش اسپینهای اتمهای اورانیوم است. باید اشاره کرد که در اتم اورانیوم تنها الکترونهای مغناطیسی، الکترونهای اوربیتال $f$ هستند، بنابراین طرز قرار گرفتن گشتاورهای مغناطیسی مربوط که با عنوان گشتاورهای $5f$ از آنها نام برده می‌شود مهمترین عامل در ایجاد گرادیان میدان الکتریکی است.

همچنین توابع عدم تقارن اربیتالهای $p$ و $d$ با استفاده از تقریبهای $GGA$ و $GGA+SO$ و $LDA+U$ در جداول ۶ و ۷ و ۸ آورده شده‌اند. هر چه مقدار عدم تقارن توزیع بار در یک اربیتال بیشتر باشد سهم آن اربیتال در گرادیان میدان الکتریکی بیشتر است. در مورد اتمهای اورانیوم عدم تقارن اربیتال $p$ بیشتر از عدم تقارن اربیتال $d$ در این اتم است، بنابراین سهم اربیتال $p$ در به وجود آوردن گرادیان میدان الکتریکی در مکان هسته $U$ بیشتر از سهم اربیتال $d$ این اتم است. با توجه به مقادیر توابع عدم تقارن اربیتالهای $p$ و $d$ و مقایسه مقادیر گرادیان میدان الکتریکی در دو فاز $GGA$ و $GGA+SO$ چنین استنباط می‌شود که تغییر حاصل شده در گرادیان میدان الکتریکی بیشتر متأثر از اربیتالهای $p$ می‌باشند و اربیتالهای $f$ در این راستا نقش بسیار جزیی دارند، چرا که در غیر اینصورت با توجه به سهمهای مشارکت اربیتالهای $f$ که به طور ناگهانی با تأثیر برهمکنش اسپین - مدار زیاد می‌شوند، می‌بایستی که مقدار $EFG$ را خیلی تغییر دهند، اما در عمل تغییر مقدار $EFG$ در مقایسه با تغییر سهم مشارکت مولفه $f-f$ خیلی کمتر است (۸/۹ - ۹/۸) که این مطلب با توجه به متفاوت بودن سهمهای مشارکت مولفه $f-f$ (سهمهای $up, dn$) توجیه می‌شود. بنابراین اربیتالهای $f$ حضور خیلی کم رنگ‌تری در مقدار گرادیان میدان الکتریکی نسبت به اربیتالهای $p$ دارند. مقایسه جداول توابع عدم تقارن در تقریبهای مختلف نشان می‌دهد که با تأثیر برهمکنش اسپین - مدار تابع عدم تقارن اربیتال $p$ نسبت به تقریب $GGA$ کاهش می‌یابد و این در حالی است که عدم تقارن اربیتال $d$ افزایش یافته است. اما

جدول ۹. سهم تماس فرمی با استفاده از تقریبهای مختلف در $USb_2$ (KGAUSS).

| | | Valance | core | $B_{hf}^{ct}$ |
|---|---|---|---|---|
| GGA+SO | U(۱) | ۹۶۸/۳۸۱ | -۹۳۵/۲۶۲ | ۳۳/۱۲۰ |
| | U(۲) | -۹۵۰/۳۲۶ | ۹۳۵/۱۳۸ | -۱۵/۱۸۹ |
| LDA+U | U(۱) | ۹۹۲/۲۲۷ | -۱۲۶۲/۴۵۳ | -۲۷۰/۲۲۶ |
| | U(۲) | -۹۸۵/۷۹۵ | ۱۲۶۱/۵۳۷ | ۲۷۵/۷۴۲ |

جدول ۱۰. سهم اربیتالی میدان مغناطیسی فوق ریز با استفاده از تقریبهای مختلف در $USb_2$ (T).

| | | up-up | dn-dn | $B_{hf}^{o}$ |
|---|---|---|---|---|
| GGA+SO | U(۱) | -۳۰۷/۱۴۶۰۲ | ۰/۶۵۰۳۵۰۰ | -۳۰۶/۴۹۵۶۷ |
| | U(۲) | -۰/۶۴۴۸۲ | ۳۰۷/۱۴۴۷۳ | ۳۰۶/۴۹۹۹۱ |
| LDA+U | U(۱) | -۳۴۴/۵۹۱۸۳ | ۰/۶۵۸۹۸۰۰ | -۳۴۳/۹۳۲۸۵ |
| | U(۲) | -۰/۶۵۷۲۰۰۰ | ۳۴۴/۸۲۹۱۸ | ۳۴۴/۱۷۱۹۹ |

جدول ۱۱. سهم اسپینی میدان مغناطیسی فوق ریز با استفاده از تقریبهای مختلف در $USb_2$ (T).

| | | up-up | dn-dn | $B_{hf}^{d}$ |
|---|---|---|---|---|
| GGA+SO | U(۱) | ۱۴/۳۴۵۷۴ | ۰/۷۴۳۵۷ | ۱۵/۰۸۹۳۱ |
| | U(۲) | -۰/۷۴۲۴۵ | -۱۴/۳۵۴۴۶ | -۱۵/۰۹۶۹۱ |
| LDA+U | U(۱) | ۱۵/۹۰۷۵۵ | ۰/۷۶۱۶۳ | ۱۶/۹۱۸ |
| | U(۲) | -۰/۷۵۹۸۸۰ | -۱۵/۹۲۹۷۴ | -۱۶/۶۸۹۶۲ |

با تأثیر برهمکنش $LDA+U$ سهم تابع عدم تقارن اربیتال $p$ باز هم کاهش می یابد در حالیکه تابع عدم تقارن اربیتال $d$ نسبت به $GGA$ افزایش و نسبت به $GGA+SO$ کاهش یافته است.

### ۶. ۲. میدان مغناطیسی فوق ریز

همچنین با توجه به علامت و مقادیر میدان مغناطیسی فوق ریز (جداول ۹-۱۱) و ممان مغناطیسی درون کرات موفین - تین اتمهای اورانیوم (MMI) (جدول ۱۳) می‌توان به دوگانگی اتمهای اورانیوم پی برد. علامت و اندازۀ این کمیتها برای اتمهای (۱) و (۴) با هم و (۲) و (۳) نیز با هم یکسان است. اما علامتهای آنها نسبت به هم قرینه می‌باشند که این مطلب مؤید یکسان نبودن اتمهای اورانیوم در این ترکیب می‌باشد. بنابراین می‌توان نتیجه گرفت برهمکنش‌های فوق ریز مورد بررسی بیشتر متأثر از ساختار مغناطیسی هستند تا ساختار بلوری. این محاسبات در دو فاز $GGA+SO$ و $LDA+U$ انجام شده‌اند. در جداول ۹، ۱۰ و ۱۱ به ترتیب داده‌های به دست آمده برای $H_{HF}^{CF}$، $H_{HF}^{O}$ و $H_{HF}^{d}$ آورده شده است. سرانجام در جدول ۱۲ میدان مغناطیسی فوق ریز کل، برای اتمهای $U$ (۱) که طبق رابطه (۱۱) مجموع مقادیر به دست آمده برای سهمهای مختلف ذکر است و با مقدار تجربی اندازه‌گیری شده مقایسه شده‌اند. توجه داریم که این داده‌ها فقط برای

جدول ۱۲. میدان مغناطیسی کل با استفاده از تقریبهای مختلف در $USb_2$ ($T$).

| | | $H_{HF} = H_{HF}^{CF} + H_{HF}^{O} + H_{HF}^{d}$ |
|---|---|---|
| exp | | ۲۷۰ ± ۲۰ |
| $GGA + SO$ | $U(۱)$ | –۲۸۸/۰۸۹ |
| | $U(۲)$ | ۲۸۹/۸۸۴۱ |
| $LDA + U$ | $U(۱)$ | –۳۵۴/۰۳۶۶ |
| | $U(۲)$ | ۳۵۵/۰۵۶۵۷ |

جدول ۱۳. ممان مغناطیسی اتمهای اورانیوم با استفاده از تقریبهای مختلف در $USb_2$.

| MMI | $U(۱)$ | $U(۲)$ | $U(۳)$ | $U(۴)$ |
|---|---|---|---|---|
| $GGA$ | ۱۷/۹۸۲۶۷ | –۱/۹۸۳۶۱ | –۱/۹۸۹۶۹ | ۱/۹۹۰۸۰ |
| $GGA + SO$ | ۱/۷۱۶۴۵ | –۱/۷۱۶۲۶ | –۱/۷۱۶۳۴ | ۱/۷۱۶۴۸ |
| $LDA + U$ | ۲/۲۸۰۱۱ | –۲/۲۸۰۱۱ | –۲/۲۷۹۶۱ | ۲/۲۷۹۴۵ |

اتمهای $U(۱), U(۲)$ (شکل ۱ را ببینید) آمده‌اند. زیرا مقادیر به دست آمده برای اورانیومهای دیگر بسیار مشابه با این مقادیر هستند. همان طور که جدول ۱۲ نشان می‌دهد نتایج اعمال برهمکنش $GGA + SO$ نسبت به برهمکنش $LDA + U$ به تجربه نزدیکتر است که می‌توان علت را به خطای ناشی از به کار نبردن مقدار دقیق پارامتر U برای این ترکیب نسبت داد.

## ۶. ۳. چگالی حالتها

سطح زیر منحنی چگالی حالتها در هر گسترهٔ انرژی برابر تعداد حالات مجاز حضور الکترون در آن گستره است. منحنیهای مربوط به کل بلور با استفاده از تقریبهای $GGA$ و $LDA + U$ در شکلهای ۴ و ۵ آورده شده‌اند.

همان طور که از این شکلها پیدا است برای کل بلور تعداد حالات با اسپین بالا برابر تعداد حالات اسپین پایین است که این امر نشانگر آنتی فرومغناطیس بودن این ترکیب است. پیش فرض تقریب $GGA$ به این صورت است که بیشتر الکترونها را در سطح فرمی قرار می‌دهد و با آنها به عنوان الکترونهای رسانش برخورد می‌کند که این مطلب در شکل مشاهده می‌شود. مشاهده می‌شود که منحنیهای $DOS$ با استفاده از تقریب $LDA + U$ و تأثیر پارامتر هابارد $U$ به عنوان یک پارامتر جایگزیده در هامیلتونی، در مقایسه با تقریب دیگر، بیشتر از سطح فرمی فاصله می‌گیرند. همچنین مقدار آن در سطح فرمی کمتر از مقدار حاصل از تقریب $GGA$ است، زیرا فرض اساسی تقریب $LDA + U$ جایگزیده بودن الکترونها و دور نمودن چگالی حالات از سطح فرمی است.

با این حال همان طور که نمودارها نشان می‌دهند، استفاده از تقریب $LDA + U$ همانند تقریب $GGA$ باز هم منحنی چگالی حالات سطح فرمی را قطع می‌کند، که بیانگر این واقعیت است که در این حالت هنوز خاصیت فلزی وجود دارد و الکترونها به طور کامل از سطح فرمی کنده نشده و گاف انرژی به وجود نیامده است. اما تأثیر جایگزیدگی این تقریب نسبت به تقریب قبلی بیشتر بوده است.

از مقایسهٔ نمودار چگالی حالات اربیتال $f$ اتم اورانیوم شکلهای ۶ و ۷ با نمودار چگالی حالات کل اتم اورانیوم شکلهای ۸ و ۹ چنین استنباط می‌شود که بیشترین سهم در چگالی حالات اتم اورانیوم مربوط به اربیتال $f$ می‌باشد. به همین دلیل رفتار و خصوصیات این ترکیب بیشتر متأثر از الکترونهای $f$ است.

همان طور که نمودارهای مربوط به اربیتال $f$ به دست آمده از اعمال تقریبهای مختلف نشان می‌دهند، شکلهای ۶ و ۷، الکترونهای این اربیتال بیشتر تمایل دارند در نزدیکی سطح فرمی قرار گیرند. علی‌رغم به کار بردن تقریب $LDA + U$ برای جایگزیده کردن این الکترونها آنها از سطح فرمی خیلی فاصله نمی‌گیرند. بنابراین می‌توان چنین برداشت نمود که

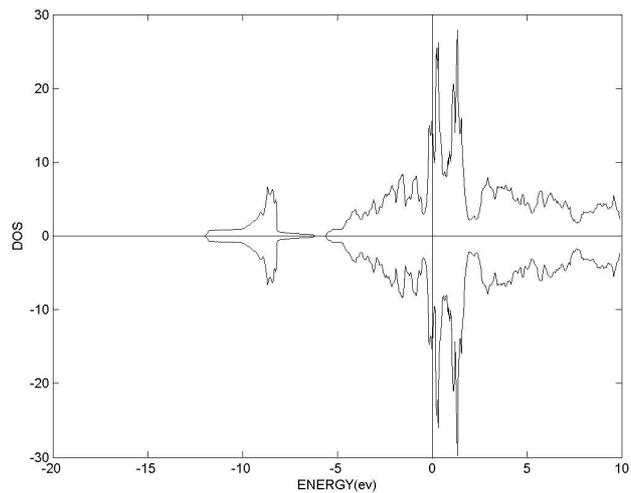

شکل ۴. DOS کل بلور $USb_2$ در فاز GGA.

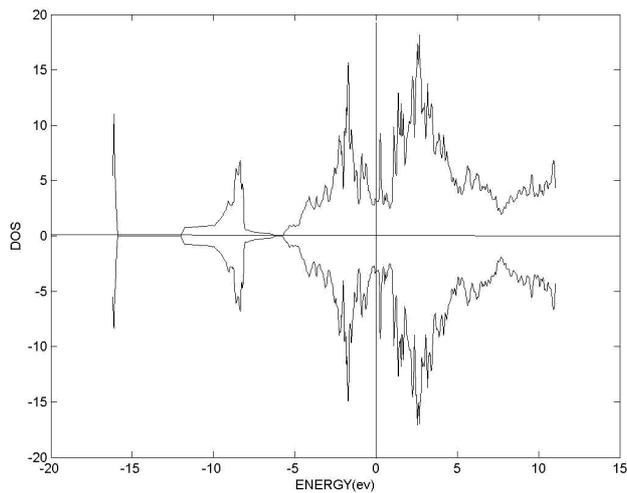

شکل ۵. DOS کل بلور $USb_2$ در فاز LDA+U.

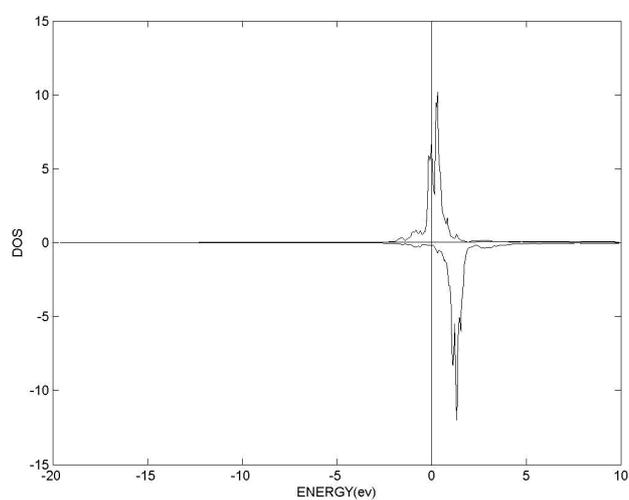

شکل ۶. DOS اربیتال f اتم اورانیوم شماره ۱۰ در فاز GGA.

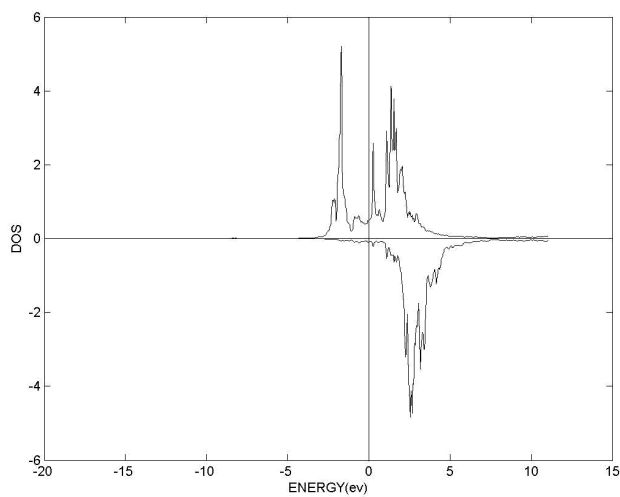

شکل ۷. DOS اربیتال f اتم اورانیوم شماره ۱۰ در فاز LDA+U.

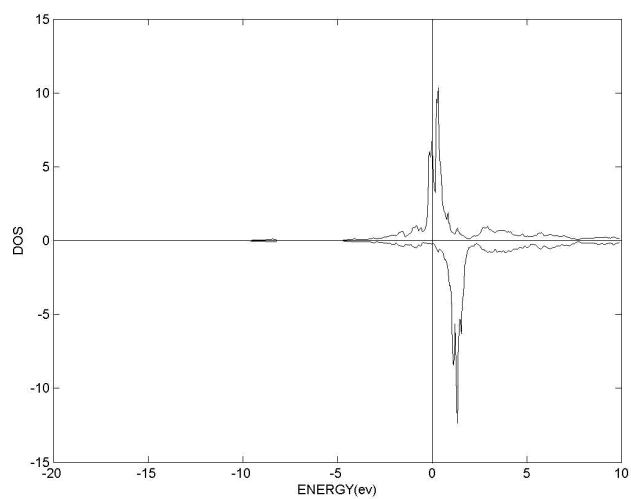

شکل ۸. DOS کل اتم اورانیوم شماره ۱۰ در فاز GGA.

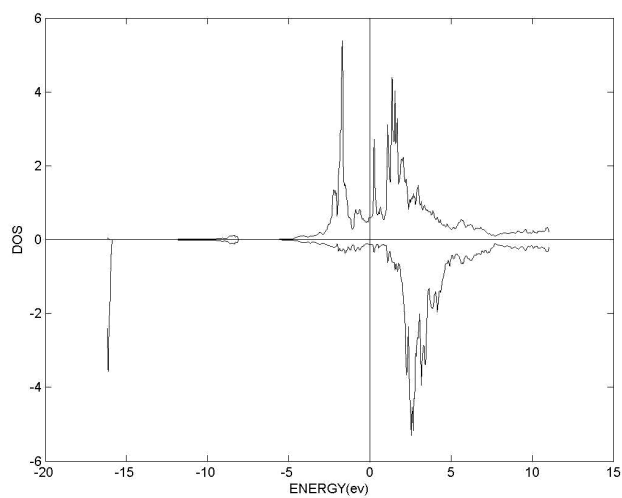

شکل ۹. DOS کل اتم اورانیوم شماره ۱۰ در فاز LDA+U.

الکترونهای $f$ در این ترکیب نه به طور کامل جایگزیده هستند و نه به طور کامل غیر جایگزیده، و در واقع رفتار دوگانه‌ای را از خود نشان می‌دهند که این مطلب در توافق با نتایج $D.Aoki$ و همکاران می‌باشد [۶]. به همین دلیل در هنگام بررسی خواص این بلور باید توجه داشته باشیم که استفاده از روشهایی که تمام الکترونها را مغزه به حساب می‌آورد (open core) جایز نیست.

## قدردانی

## مراجع